\documentclass[journal, twocolumn]{IEEEtran}
\usepackage{graphicx}
\usepackage{amssymb}
\usepackage[dvipsnames]{xcolor}
\usepackage{subcaption}
\usepackage{amsmath}
\usepackage{dblfloatfix}
\usepackage{authblk}
\usepackage{float}
\newcommand\size{0.9}

\begin{document}

\title{Physics-enhanced machine learning for virtual fluorescence microscopy}

\author[1*]{Colin L. Cooke\thanks{*Colin Cooke is corresponding author: colin.cooke@duke.edu}\thanks{Source code available: www.github.com/clvcooke/virtual-fluorescence}}
\author[1]{Fanjie Kong}
\author[1]{Amey Chaware}
\author[2]{Kevin C. Zhou}
\author[1]{Kanghyun Kim}
\author[1]{Rong Xu}
\author[3]{D. Michael Ando}
\author[3]{Samuel J. Yang}
\author[2]{Pavan Chandra Konda}
\author[1,2]{Roarke Horstmeyer}
\affil[1]{Duke University, Electrical and Computer Engineering Department}
\affil[2]{Duke University, Biomedical Engineering Department}
\affil[3]{Google Research, Applied Science Team}

\maketitle

\begin{abstract}
This paper introduces a new method of data-driven microscope design for virtual fluorescence microscopy. Our results show that by including a model of illumination within the first layers of a deep convolutional neural network, it is possible to learn task-specific LED patterns that substantially improve the ability to infer fluorescence image information from unstained transmission microscopy images. We validated our method on two different experimental setups, with different magnifications and different sample types, to show a consistent improvement in performance as compared to conventional illumination methods. Additionally, to understand the importance of learned illumination on inference task, we varied the dynamic range of the fluorescent image targets (from one to seven bits), and showed that the margin of improvement for learned patterns increased with the information content of the target. This work demonstrates the power of programmable optical elements at enabling better machine learning algorithm performance and at providing physical insight into next generation of machine-controlled imaging systems.
\end{abstract}
 
\section{Introduction}
\begin{figure*}[tb]
    \centering
    \includegraphics[width=0.9\textwidth]{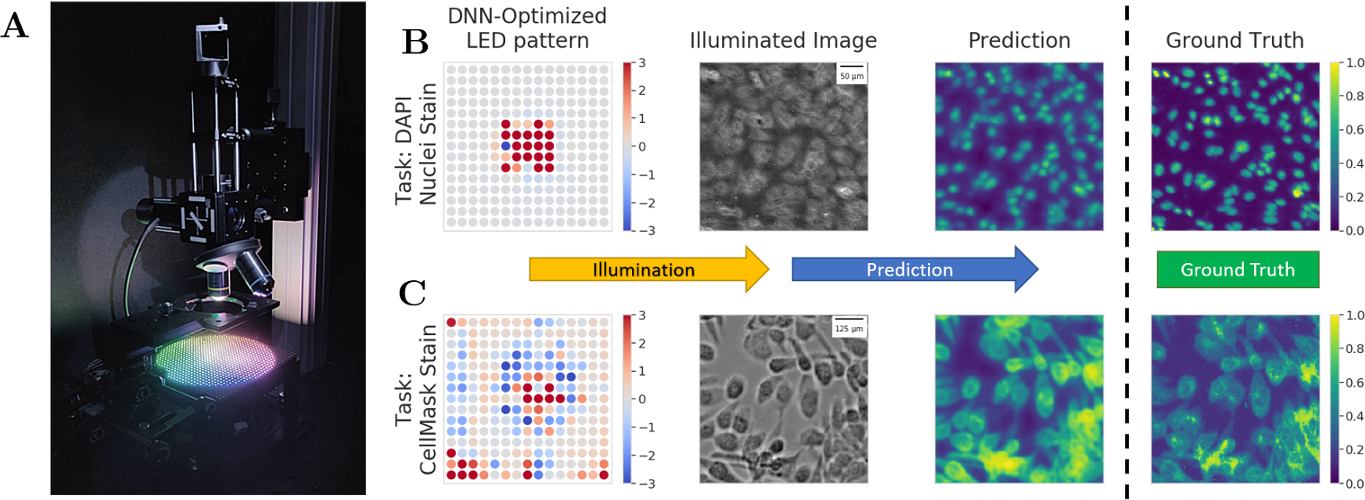}
    \caption{\textbf{(A)} Experimental setup: Microscope with both fluorescence and non-fluorescence imaging paths centered above a programmable LED array for training data capture. \textbf{(B/C)} Example fluorescent inference pipelines: The optimized LED pattern illuminates a sample to form an image. The neural network processes LED-illuminated images to produce an estimated fluorescent result. The ground truth is shown to the right of the inference result. Example B shows the \textit{HeLa} task while Example C shows the \textit{PAN} task.}
    \label{fig:image_pipeline}
\end{figure*}
The optical microscope remains a critical tool across a wide variety of disciplines. Examples include high-content screening in biology labs, quality control and defect detection in factories, and automated digitization of pathology slides in clinics. With the continued growth of automated software analysis tools, many microscope images are now rarely viewed directly in their raw format by humans, but are instead commonly processed by a computer first. Examples include the automatic classification of different cell types within large cell cultures \cite{van2016deep}, segmentation of cancerous  areas from thin pathology tissue sections \cite{xu2017deep}, and, as focused upon here, the automatic creation of fluorescent images from bright-field data~\cite{christiansen2018silico}.  

Despite the continued rapid development of automated image analysis software, the microscope’s hardware has changed relatively little over the past several centuries.  Most current microscopes still consist of standard illumination units and objective lenses that are optimized for direct human inspection. The physics of optical microscopes enforces several physical limitations, including a limited resolution, field-of-view, image contrast, and depth-of-field, for example, which restrict the amount of information that can be captured within each image. The standard design of a microscope biases this limited information towards human analysis, potentially impacting the accuracy of automated analysis.

Here, we attempt to optimize the hardware of a new microscope design to improve the accuracy of automated image processing by a deep neural network (DNN). Our aim in this work is to establish optimal hardware settings to improve the particular DNN task of image inference. To achieve this goal, we present a modified learning network that includes a physical model of our experimental microscope, which we jointly optimize during DNN training. In this work, we limit our physical model to include only the spectral and angular properties of the microscope's illumination, provided by a programmable LED array, but leave open the possibility of also considering many other important parameters (focus setting, lens design, detector properties) in future work. Our proposed network models the microscope illumination pattern as a set of linear weights that are directly integrated into the DNN, allowing the calculation of gradients through back-propagation and end-to-end optimization during supervised training. After training, the optimized ``physical" weights can be interpreted as the distribution of optimal LED brightnesses and colors to use in our experimental imaging setup, which transfers performance gains seen in training to a physical setup.

For the specific goal of DNN-optimized microscope illumination for image-to-image inference, we aim here to train a DNN to convert optimally illuminated microscope imagery into data obtained from a simultaneously captured fluorescence image. This goal of bright-field to fluorescence image inference, also termed ``in-silico labeling"\cite{christiansen2018silico}, has recently received interest as a promising means to avoid the need to fluorescently label specimens and to instead simply rely on post-processing standard bright-field image data. Prior work with this effort has typically required a relatively large data overhead (i.e., acquisition of 10 or more bright-field images per inference task), and has offered a limited amount of physical insight into performance variations as a function of collected data.

In this work, we use a DNN to jointly optimize the illumination to both reduce the number of required images for accurate fluorescence image inference, and to explore how performance scales with the amount of inferred information. To achieve this latter goal, we vary the precision of the desired fluorescent output to consider a range of inference tasks, from binary image segmentation (one bit) to the prediction of complete fluorescent images (seven bits). For each level of precision, we examine how the converged LED pattern changes, and how these patterns trend with precision. The optimized patterns not only yield higher accuracy results, but also provide a certain degree of physical intuition between scattered bright-field light and fluorescent emission that can be used to improve future data collection strategies.

\section{Related work}
\begin{figure*}[tb]
    \centering
    \includegraphics[width=1.5\columnwidth]{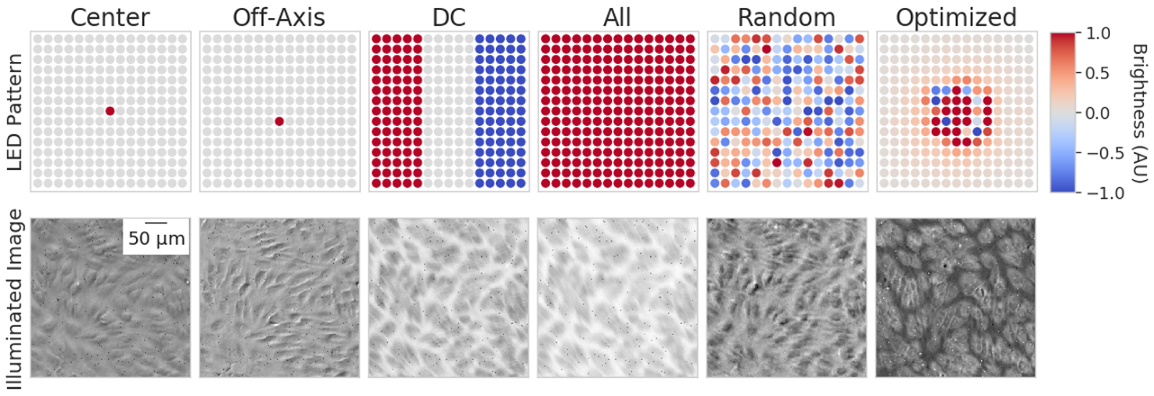}
    \caption{Standard LED patterns tested in this work, and resultant illuminated images}
    \label{fig:standard_patterns}
\end{figure*}
\begin{figure}[tb]
    \centering
    \includegraphics[width=\columnwidth]{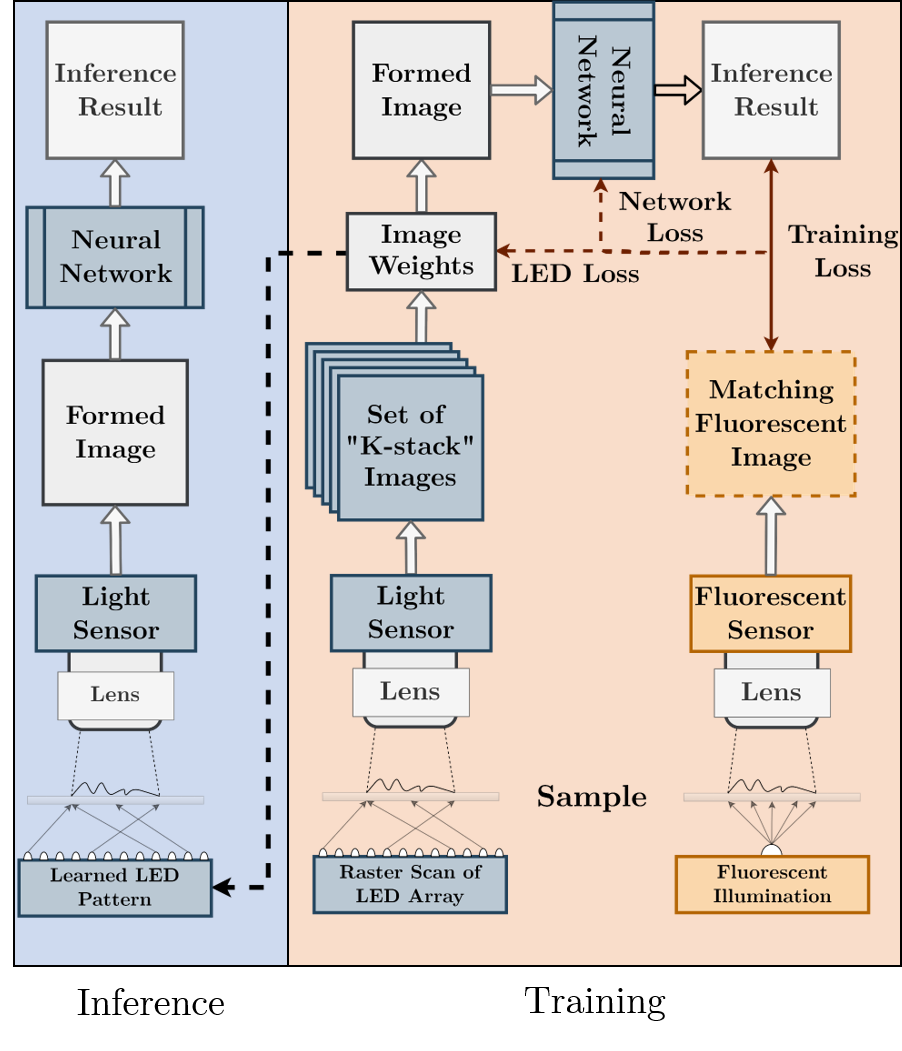}
    \caption{Diagram of  training and inference pipelines for physics-enhanced fluorescence image inference. The orange region indicates the components involved in training, while the blue region contains components used for inference. The fluorescence image and LED-generated "k-stack" images are captured through the same objective lens and split via a beamsplitter.  Multiplicative weights for the LED images are learned during training are transformed into LED brightnesses during fluorescence image inference.}
    \label{fig:training_setup}
\end{figure}
In recent years, convolutional neural networks (CNNs) have become commonplace for both medical and natural image processing \cite{geert}. Segmenting images to find specific cell types or sub-cellular features (e.g. cell nuclei), for example, is now a common biomedical image analysis task that CNNs excel at. The U-net structure \cite{unet}, perhaps one of the most widely used CNN architectures, has been applied across a wide variety of segmentation tasks \cite{Unet_perf} and makes efficient use of annotated data during training.

As the use of neural networks continues to increase in popularity, many researchers are now also applying them to automatically analyze fluorescent imagery. Belthangady et al. \cite{belthangady2019applications} recently reviewed this increasing body of work and summarized it into two general categories: virtual staining and fluorescent image enhancement. 

``In-silico" labelling is the process of using bright-field images to predict fluorescent images. While still a relatively new concept, several recent works demonstrate that DNNs can be quite effective at this task, which suggests that certain future experiments may forgo fluorescent imaging and staining entirely. Christiansen et al. \cite{christiansen2018silico} first demonstrated this concept by developing a CNN which predicted seven distinct fluorescent channels from a set of several dozen uniquely focused bright-field images. Ounkomol et al. \cite{ounkomol2018label} then used a modified U-net structure to predict 3D fluorescent images using 3D bright-field data. Chen et al. subsequently \cite{chen2019augmented} applied this technology through the design of an \textit{Augmented Reality Microscope} which overlays a neural network's predictions on top of the underlying bright-field data. In a related effort, Rivenson et al. \cite{rivenson2019virtual} used a neural network to infer histologically stained images from unstained tissue with high accuracy and visual quality. These works all demonstrate that it is possible to infer information revealed by a fluorescent or histological stain from unmodified bright-field data, albeit at varying levels of accuracy. 

Fluorescent image enhancement focuses on improving the quality of existing fluorescent images. Weigert et al. \cite{weigert2018content} developed a content-aware image restoration method powered by a CNN. This work showed that fluorescent image restoration was possible by predicting high resolution fluorescent images from ones which were under-sampled. In an earlier work, Weigert et al. \cite{weigert2017isotropic} used a CNN to perform isotropic reconstruction of 3D fluorescent data. Through these works, we can infer that data contained within fluorescent images is potentially redundant, and through the assumption of key underlying features, it is at times possible to enhance image quality via CNN post-processing.   

However, in most studies, the focus is on processing data that has already been captured, rather than attempting to influence or improve the image acquisition process. While an early work used simple neural networks to effectively design components of optical systems \cite{macd}, the first work (to the best of our knowledge) to examine hardware optimization in the context of CNNs was by Chakrabarti \cite{chakrabarti}, who presented an optimal pixel level color-filter layout for color image reconstruction. A number of subsequent works have considered how to merge the optimization of various imaging hardware components into a differentiable optimization network \cite{dirtypixels,recon, sitzmann2018end, chang2018hybrid, hershko2019multicolor, xue2019reliable,del2020learned, sun2020learning, barbastathis2019use}. However, few of these works have proposed the use of CNNs to optimize the image capture process for automated decision tasks (e.g., classification, object detection, image segmentation, etc.). 

Such an approach was recently considered by Horstmeyer et al. \cite{roarke}, who suggested the use of a "physical layer" in a CNN to identify optimized illumination patterns that can improve automated classification accuracy of the malaria infected blood cell detection. Other related work has used DNNs to design custom optical elements to improve an image's depth-of-field~\cite{gordon}, to inform new types of illumination for improved phase contrast imaging \cite{waller,rainer}, achieve superior resolution \cite{ganapati,zheng}, or infer color from a gray-scale image \cite{hershko2019multicolor} . The goal of these works was not to improve the accuracy of automated image inference, as considered here.

In terms of using a microscope's illumination to achieve new functionalities, prior work has clearly shown the benefits of applying programmable LED array lighting. This includes variable bright-field and dark-field imaging \cite{zheng2011microscopy}, measurement of a specimen's surface gradient\cite{waller,rainer}, and quantitative phase imaging \cite{ou2013quantitative} to name a few. Mathematically rigorous methods have also been used to combine variably-illuminated images to increase image resolution. Two prominent examples are Fourier ptychographic microscopy (FPM) ~\cite{zheng2013wide} and structured illumination microscopy (SIM) \cite{gustafsson2000surpassing}. These works highlight the benefit of controlling illumination and provide ample evidence that it should be targeted for optimization.

\section{Methods}

\label{sec:methods}
\subsection{Image Formation}
In this work, we collected and processed data across two different experimental setups. Both setups used an off-the-shelf LED array (\textit{Adafruit} product ID 607), which provided the capability to turn on specific LED array patterns at different colors (e.g., red, green and blue) for multi-angle and multi-spectral specimen illumination (Figure \ref{fig:image_pipeline}A). In each setup, the LED array was placed sufficiently far from the sample plane such that the illumination from each LED could be modelled as a plane wave at the sample, propagating at a unique angle corresponding to its relative position. This type of illumination has been previously used for FPM \cite{zheng2013wide, konda2020fourier}, as well as phase contrast \cite{waller,rainer} and super-resolved 3D imaging \cite{3dfpm, tian20153d, zhou2019diffraction}, among other enhancement techniques. As noted above, illumination from a wide variety of angles and colors provides diverse information about biological specimens, which are primarily transparent, that is not available under normal illumination (see Figure \ref{fig:standard_patterns}). However, it is not directly obvious which type of illumination is best-suited for mapping non-fluorescent image data into fluorescence stained image data. The optimal patterns will likely depend upon which sample features are fluorescently labeled, as well as properties of the specimen's complex index of refraction. In this work, we consider two separate fluorescent labeled sample categories - one in which the cell nucleus is labeled, and a second in which the cell membrane is labeled - to verify this hypothesis. In addition, the specific inference task may also impact the optimal illumination distributions (i.e., segmentation versus image formation) in a non-obvious way. To solve for such task specific illumination patterns, we propose to use a modified DNN, which includes a "physical layer"~\cite{roarke} for joint hardware optimization.

In this type of approach, the supervised machine learning network itself jointly determines the optimal distribution of LEDs' brightness and colors to use for sample illumination, and to perform the subsequent image-to-image inference task. As we describe below, the process of LED illumination can be modeled by a single physical layer, which we prepend to the front of a DNN U-Net architecture for improved image-to-image inference. After the network training is complete, the optimized weights within the physical layer (in this case, representing the angular and spectral distribution of the sample illumination) inform us of a better optical design for each specific inference task at hand. Figure \ref{fig:training_setup} shows how this process is realized for both training and inference.

To mathematically model image formation, we express the $j$th specimen of interest as the complex function $o_j(r,\lambda)$. Given $N$ LEDs within the array, we can denote the amplitude of the plane wave emitted by each colored LED emitting at center wavelength $\lambda$ as a weight $\sqrt{w_n(\lambda)}$ for $n = 1, \dots, N$. Each LED in the array illuminates the sample with a coherent plane wave from a unique angle $\theta_n$. Since, the LEDs are mutually incoherent with respect to each other, the image formed by an illumination composed of multiple LEDs is equivalent to the incoherent sum of images obtained by illuminating the member LEDs individually. Assuming a thin specimen, for example, we may write the $j$th detected quasi-monochromatic image $I_j^\prime$ as the incoherent sum, 
\begin{equation}
\label{eq:image_form}
    I_j^\prime(r) = \sum_{\lambda \in \{RGB\}} \sum_{n=1}^N |o_j(r,\lambda) \cdot \sqrt{w_n(\lambda)}e^{i\mathbf{k}_n\cdot\mathbf{r}} \star h(r)|^2
\end{equation} 
where, $\sqrt{w_n(\lambda)}e^{i\mathbf{k}_n\cdot\mathbf{r}}$ describes the plane wave generated by the $n$th LED with intensity $w_n(\lambda)$ at wavelength $\lambda$ across the sample plane with coordinate $\mathbf{r}$, $\mathbf{k}_n=\frac{2\pi}{\lambda}\sin{\theta_n}$ denotes the plane wave transverse wavevector with respect to the optical axis, and $h(r)$ denotes the microscope's coherent point-spread function and describes the imaging system blur.

As the LED brightness $w_n(\lambda)$ is a scalar quantity, we can factor it out of the summation in Eq. \ref{eq:image_form}. Furthermore, if we denote the image of the $j$th sample formed when it is illuminated by the $n$th LED at a fixed brightness and wavelength as $I_{j,n}(\lambda) =  |o_j(\lambda) \cdot e^{i\mathbf{k}_n\cdot\mathbf{r}} \star h|^2$,
then we arrive at a simple linear model for image formation for the $j$th sample:
\begin{equation}
I_j^\prime(r) = \sum_{\lambda} \sum_{n=1}^N w_n(\lambda)I_{j,n}(r,\lambda).
\label{eq:phys_layer}
\end{equation}
The detected image under illumination from a particular LED pattern, where each LED has a brightness $w_n(\lambda)$, is equal to the weighted sum of images captured by turning on each LED individually. The detected image $I_j^\prime$ is then entered into the neural network for processing. Equation \ref{eq:phys_layer} represents our physical layer for microscope illumination, and the fully differentiable optimization pipeline is shown in Figure \ref{fig:training_setup}.

To find the set of weights $w_n(\lambda)$ which best parameterize the transform in Eq. \ref{eq:phys_layer} for the subsequent inference task, we must have access to all $N$ uniquely illuminated images $I_{j,n}$ during network training. However, once training is complete, the set of optimized weights $\{w_n\}$ are then mapped onto the physical LED matrix to allow for acquisition of a single optimally illuminated image, which is then processed by the remainder of the DNN for the fluorescent image inference task.

\subsection{Network Design}
We used a consistent U-Net architecture that we prepended with a physical layer to model Equation.~\ref{eq:phys_layer}. The exact architecture and layer configurations, used in all the reported experiments, is detailed  in Appendix \ref{app:nn_config}. The weights within the physical layer were unconstrained, meaning they could take on both positive and negative values. To experimentally realize an unconstrained set of illumination weights $\mathbf{w}$, we captured two images (instead of one) - a first  with the positive set of weights, and a second with the negative set, before subtracting the second image from the first for the final result.

Although the experimentally captured measurements inherently contain Poisson noise, the digital simulation of multi-LED images used during training will have an artifactually inflated SNR (e.g., averaging $N$ images produces a $\sqrt{N}$ improvement in SNR).
To compensate for this we introduced a \textit{Noise Layer} which adds dynamically generated Gaussian random noise to the data after the first physical layer:
\begin{equation}
     I_j^{\prime\prime} = \mathcal{N}(\mu=I_j^\prime, \sigma^2=k\times |I_j^\prime|)
     \label{eq:noise_model}
\end{equation}
Additive noise is modelled on a per-pixel level with Equation \ref{eq:noise_model}, using a hyperparameter $k$ to control the scale of the noise in proportion to the pixel intensity. Note that the variance of the random noise is proportional to the image pixel intensity itself, and thus is consistent with a Poisson noise model.

Finally, an $L1$ penalty within the network cost function (absolute deviation from zero) was applied to the physical layer weights $w_n$. This $L1$ penalty term is given a small weight proportional to the magnitude of the gradients to drive weights to zero, if and only if they are not significantly contributing to the task performance (see details in Appendix \ref{app:nn_config}). The addition of this type of penalty reduces variance across random seeds and aids in interpretation of the resulting LED patterns. 

\section{Experiments}

We examined two types of fluorescently labeled biological specimen in the following experiments. For each experiment, a square $15\times15$ grid of multi-color LEDs was used for illumination, where each multi-color LED included 3 spectral channels (with center wavelengths $\lambda=$480, 540, and 632 nm), representing $N=15\times 15\times 3 = 675$ different illumination sources. For each imaging experiment, an LED image stack was acquired by turning on each LED individually and capturing a unique image. To provide target labels, each sample was also illuminated with a UV fluorescent excitation source and captured via a separate path containing a fluorescent emission filter, but using the same optical imaging setup that collects the bright-field image data. By using the same optical setup to capture the same FOV, we ensured that sample positioning and lens distortion factors remained constant across bright-field and fluorescent channels. Multiple non-overlapping fields-of-view were captured for each specimen of interest.

After data capture, each full field-of-view dataset was split into $256\times256$ pixel sections, with each bright-field section having $675$ channels (one channel for each LED configuration). These formed the $256\times256\times675$ datacubes, which were fed into our modified DNN for network training. The matching fluorescent channel was used as a target label for training, and is not present during inference. After training, the network produced a 675 element optimizable LED weight vector $\mathbf{w}$. To compare our optimized illumination results with other image collection approaches, we ran the same training process using a fixed $\mathbf{w}$ (i.e., without optimizing the physical layer), where the values correspond to common illumination configurations often seen in microscopy \cite{tian2015quantitative}:
\begin{enumerate}
    \item \emph{Center} is spatially coherent illumination from only the center LED and 3 colors
    \item \emph{All} is spatially incoherent illumination from all 225 LEDs and 3 colors
    \item \emph{DC} stands for the differential contrast illumination method
    \item \emph{Off-axis} is from an LED located 4 mm off the optical axis, illuminating at approximately $3^\circ$
    \item \emph{Random} is a randomly selected set of LED brightness values
\end{enumerate}

For each configuration and examined dataset, the neural network architecture remained unchanged and was trained using the same hyperparameters.
\begin{figure}[tb]
    \centering
    \includegraphics[width=\columnwidth]{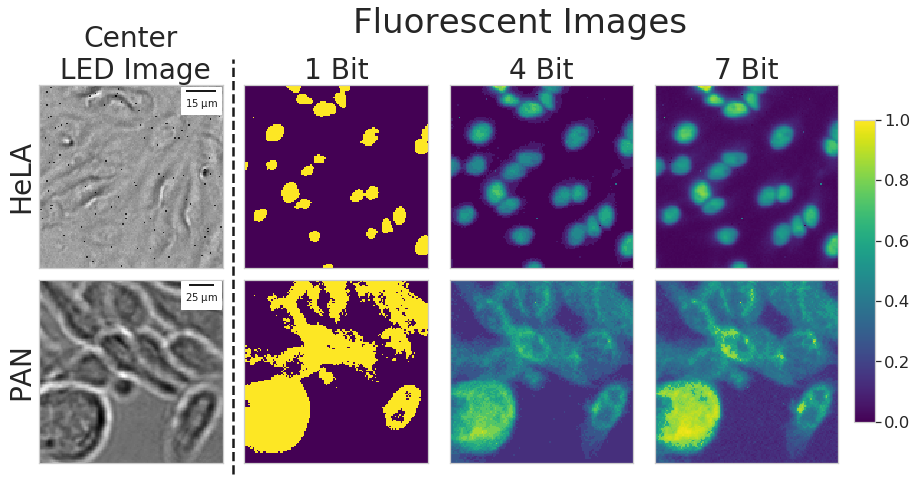}
    \caption{Examples of inferred fluorescent images at varying bit depths. Moving from left (1-bit segmentation) to right (7-bit image) we observe that not only does increasing the bit depth of the k-means quantized image change the amount of information that the network must reconstruct, but also changes where in the intensity range various features occur.}
    \label{fig:example_io}
\end{figure}
\subsection{Data Label Generation}
To understand the impact of the optimized illumination pattern $\mathbf{w}$, we varied the task difficulty by modifying the fidelity of the target label. The 7 bit ground truth fluorescence image was converted to 7 different images with different bit depths, ranging from 1 bit to 7 bits of dynamic range. Here, we hypothesize that reconstruction of a 1-bit fluorescence image, which is effectively a segmentation mask, is easier than accurately recovering a 7 bit fluorescence image. It should be noted that the input bright-field images (the training data) remains same for all tested cases. This allowed the same data to be used to train a neural network to perform both binary image segmentation task (1 bit precision), and to separately train another network to perform a full fluorescent image inference task (7 bit precision), and to also consider all tasks that span these two common efforts. Figure \ref{fig:example_io} shows how by varying the precision of the desired network output, it is possible to vary the difficulty of the subsequent image inference task, from producing a prediction that contains $1$ bits to $7$ bits per pixel. 

To achieve a balance of pixel values across each discretized histogram within the label images, we employed a global k-means quantization strategy, where $k=2^{bits}$. To find the mean values a naive k-means algorithm was applied on a flattened version of the entire dataset, treating each pixel value as an independent value. The initial mean values were uniformly distributed across the space between zero and one: $m_i = \{ \frac{i}{k}| 1 \leq i \leq k\}$. Mean values were iteratively adjusted until either all values had converged (with a threshold of: $1\mathrm{e}{-05}$) or $20$ iterations passed.

The inference results were only rounded to meet the appropriate precision of the target bit level, with no further post processing (k-means or otherwise).

\subsection{Cell Nuclei}
\label{sec:hela}
The data used in the first experiment was originally acquired by a FPM microscope \cite{data} and contains 90\% confluent HeLa cells, stained with a fluorescent nuclear label (DAPI). The employed microscope included a two-lens arrangement with an f = 200 mm tube lens (ITL200, Thorlabs) and an f = 50 mm Nikon lens (f/1.8D AF Nikkor). This two-lens setup has a collection numerical aperture (NA) of 0.085 with 3.87x magnification. The sample was placed at the front focal plane of the f = 50 mm lens and a CCD detector (pixel size 5.5 $\mu$m, Prosilica GX6600) captured images of the sample under LED matrix illumination (one LED at a time, as noted above). The LED array was placed 80 mm behind the sample with 32x32 individually addressable elements (pitch size 4 mm), of which only the inner 15x15 were utilized. Full-FOV images from all illumination angles and colors were divided into 442 unique $256\times256\times675$ datacubes. The set of datacubes was randomly split into training, testing, and validation sets containing 356, 48, and 48 images respectively.
\begin{figure*}[tb]
    \centering
    \begin{subfigure}[t]{\columnwidth}
        \centering
        \includegraphics[width=\textwidth]{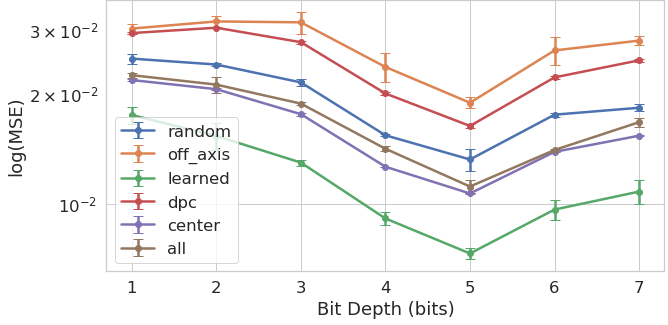}
        \caption{HeLa task: MSE of test set inference results}
        \label{fig:hela_perf}
    \end{subfigure}%
    \hfill
    \begin{subfigure}[t]{\columnwidth}
        \centering
        \includegraphics[width=\textwidth]{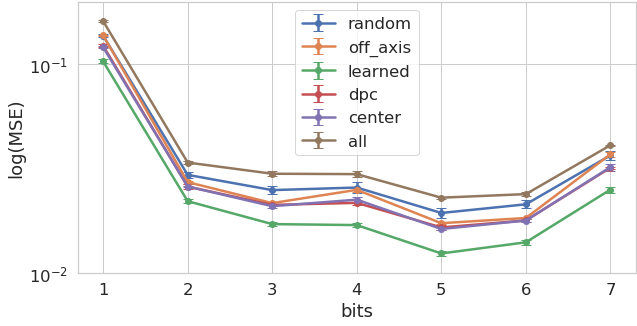}
        \caption{PAN task: MSE of test set inference results}
        \label{fig:pan_perf}
    \end{subfigure}
    \begin{subfigure}[t]{\columnwidth}
        \centering
        \includegraphics[width=\textwidth]{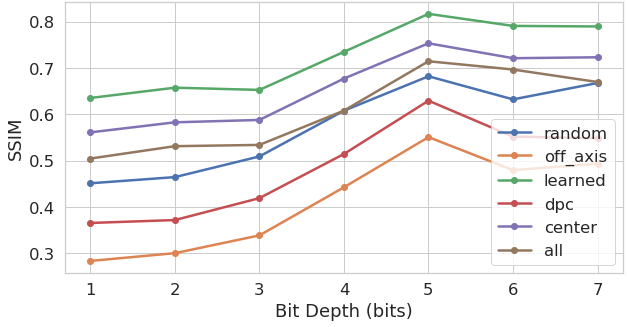}
        \caption{HeLa task: SSIM of inference results compared to fluorescent labels}
        \label{fig:hela_ssim}
    \end{subfigure}
    \hfill 
    \begin{subfigure}[t]{\columnwidth}
        \centering
        \includegraphics[width=\textwidth]{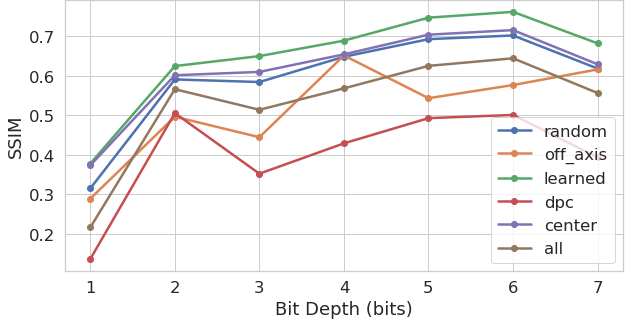}
        \caption{PAN task: SSIM of inference results compared to fluorescent labels}
        \label{fig:pan_ssim}
    \end{subfigure}
    \caption{Performance statistics of both tasks across all configurations showed that learned illumination consistently improves inference results. MSE was the target statistic to measure performance during training, while SSIM was used as an alternative measure to compare image inference quality. All statistics are averaged across three random seeds.}
    \label{fig:perf}%
\end{figure*}
\subsection{Cell Membrane}
For the second set of experiments, we captured images of Pan 16 pancreatic cancer cells stained using CellMask Green plasma membrane stain (C37608). Similarly to the dataset described in Section \ref{sec:hela}, an identical LED array was placed 60mm beneath the biological sample and of which the inner 15x15 grid was used. An Olympus PlanN 0.25NA 10x objective lens was used in conjunction with a Basler Ace (acA4024-29um) CMOS sensor. To capture the matching fluorescence data a set of Thorlabs filters (MDF-GFP - GFP Excitation, Emission, and Dichroic Filters) were inserted into the optical setup. Figure \ref{fig:image_pipeline} shows a photo of the setup used. The data was split into test/train/validation sets such that data from each sample only existed in one of the sets, providing a robust means of gauging generalization and overfitting. When split into $256\times256$ images there were 820/108/108 samples in the train, validation and test sets respectively. 

\section{Results and Discussion}
\subsection{Performance}
Across both tasks and all bit depth configurations, we found that the optimized illumination patterns determined by our physical layer outperformed all other illumination arrangements (coherent and incoherent bright-field, dark-field, random and phase contrast). Figure \ref{fig:perf} plots the mean squared error (MSE) of each inference task as a function of inference bit depth (i.e., inference difficulty), cell type, and physical layer parameterization. This main result shows that jointly optimized microscope hardware can lead to superior image inference performance across a wide range of tasks - ranging from image segmentation to  virtual fluorescence imaging.

Although MSE provides an accurate measure of the task performance, it is less suited to understand the perceptual quality of the results. We thus also computed the structural similarity index (SSIM) \cite{wang2004image} across the test set for both tasks (results shown in Figure \ref{fig:perf}). Here we can see that the reduction in mean squared error brought by the optimized illumination carries onto a higher perceptual quality. 
\begin{figure}[tb]
    \centering
    \includegraphics[width=\columnwidth]{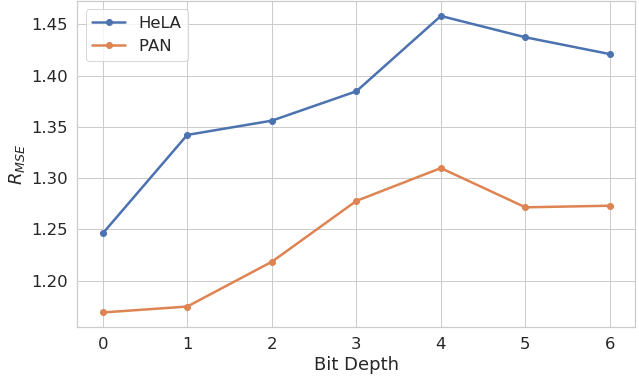}
    \caption{Comparing relative performance across tasks and bit depths using the best of standard illumination patterns compared to the learned patterns. A trend of down and to the right indicates greater performance benefits at higher bit depths.}
    \label{fig:relative_perf}
\end{figure}

\subsection{Performance versus inference task}
\label{sec:perf_vs_depth}
By varying the bit depth of the output label, we were able to test our approach with seven different configurations. We first observe that the gap in performance (both MSE and SSIM -- Figure \ref{fig:perf}) between the inference results created by the optimized LED pattern ("learned"), and the alternative standard LED illumination patterns, is quite consistent. Given that the pixel values within all of the output labels are normalized to  a $[0,1]$ range, this result is somewhat consistent with our expectations. The increase in difficulty of the task is partially offset through the decrease in the "cost" of being wrong. 

To get a better understanding of the interaction of task difficulty and performance we examined the relative performance of the "learned" method compared to the best of the "standard" methods. To do this we constructed a relative performance metric, $R_{MSE}$, which is the quotient of the minimum MSE among the standard patterns' MSE and the learned configurations MSE, for each bit depth $K$. See equations \ref{eq:mse_standard} and \ref{eq:relative_mse}.
\begin{align}
    \label{eq:mse_standard}
    MSE^K_{standard} &= min(MSE^K_{cent.,all,oa,rand.,dcp})\\
    \label{eq:relative_mse}
    R^K_{MSE} &= \frac{ MSE^K_{standard}}{MSE^K_{learned}}
\end{align}

Within Figure \ref{fig:relative_perf} we show that the relative performance of the learned LED configuration improves with difficulty when compared to the best standard configuration. This illustrates how the joint optimization process provides a consistent performance improvement that becomes more important as the difficulty of the task increases. 

\begin{figure}
    \centering
    \includegraphics[width=\columnwidth]{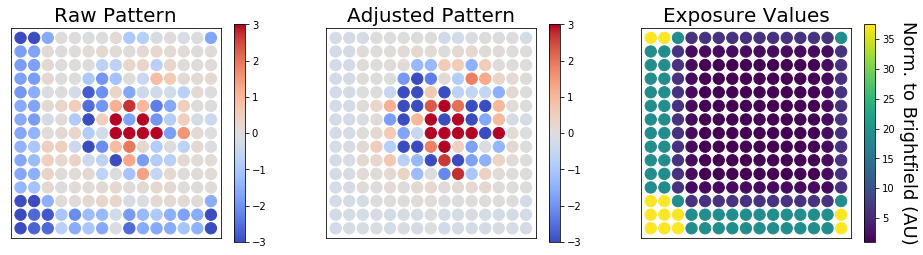}
    \caption{Exposure adjustment for visualization: For the PAN task a non-uniform exposure was used during data collection, to enable accurate visualization the PAN LED patterns were normalized by their per-LED exposure value}
    \label{fig:exp_adjustment}
\end{figure}

\begin{figure}[tb]
    \centering
    \begin{subfigure}[t]{\columnwidth}
        \centering
        \includegraphics[width=\textwidth]{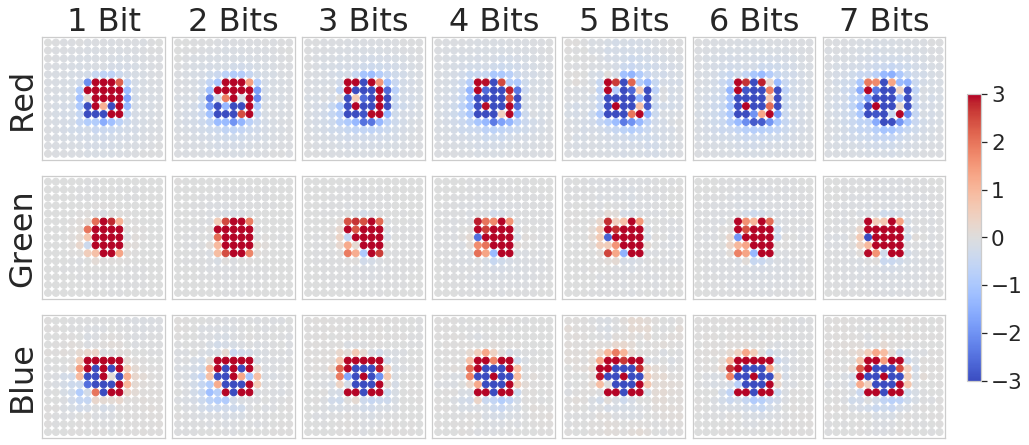}
        \caption{HeLa cell task}
        \label{fig:hela_patterns}
    \end{subfigure}%
     \\
    \begin{subfigure}[t]{\columnwidth}
        \centering
        \includegraphics[width=\textwidth]{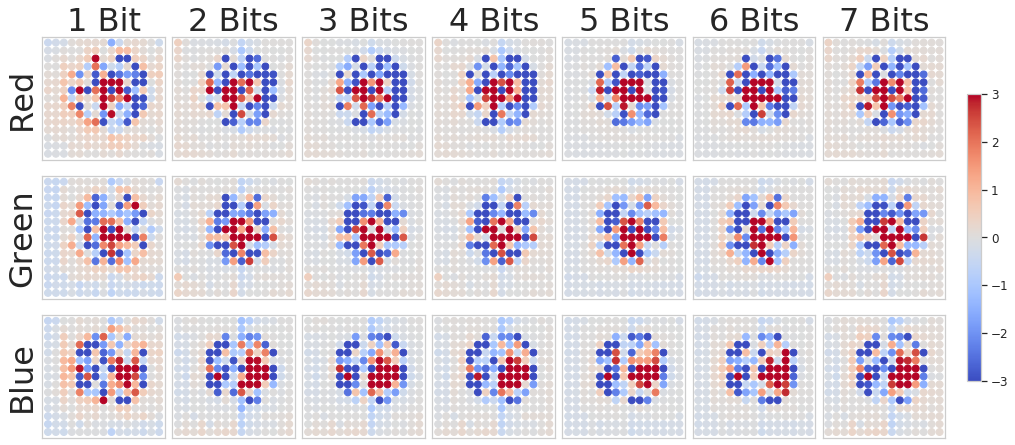}
        \caption{Pancreatic cancer cell task.}
        \label{fig:pan_patterns}
    \end{subfigure}
    \caption{LED Patterns learned through joint optimization: the patterns are shown for each bit depth configuration. The color channel indicated at the top of each row is the color of the LEDs which were illuminated. The colorbar indicates the brightness that each LED was configured to.}
    \label{fig:optimized_patterns}
\end{figure}

\subsection{LED Parameterization}
In addition to improved performance, the DNN-optimized LED illumination patterns also offer several interesting physical insights. To aid in interpretation of the patterns, each LED was normalized by it's exposure (as shown in Figure \ref{fig:exp_adjustment}), as using different exposures per LED enabled us to fully utilize the dynamic range of our image sensor. Figure \ref{fig:optimized_patterns} displays the learned patterns as a function of inference task (1-bit to 7-bit) on a per-color-channel basis for both tested specimens. 

These illumination patterns highlight the importance of two primary features for transmitting visible light information that is most relevant to a particular fluorescent specimen and stain. First, producing phase contrast by illuminating differently within the bright-field (the inner 3x3 LEDs for the HeLa task, and the inner 5x5 LEDs for the PAN task) and dark-field appears important across all tasks. Second, providing color contrast (e.g., green for the positive image and red for the negative image) also appears important, most notably for the HeLa nuclei prediction task. The pattern differences on a per-task (HeLa vs. PAN) can be attributed both to the differences between setups (notably that the PAN task setup included more bright-field LEDs) as well as the differences between samples. It should be noted that for the PAN task the LED patterns appear shifted, this is due to an indexing error which occurred during the data-preprocessing steps and does not hold any physical significance.

\begin{figure}[htb]
    \centering
    \begin{subfigure}[t]{\columnwidth}
        \centering
        \includegraphics[width=\textwidth]{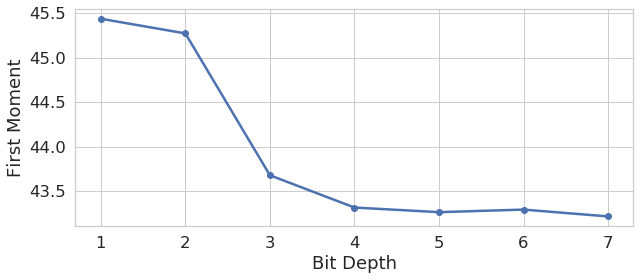}
        \caption{HeLa Task}
        \label{fig:freq_dist_hela}
    \end{subfigure}%
    \\ 
    \begin{subfigure}[t]{\columnwidth}
        \centering
        \includegraphics[width=\textwidth]{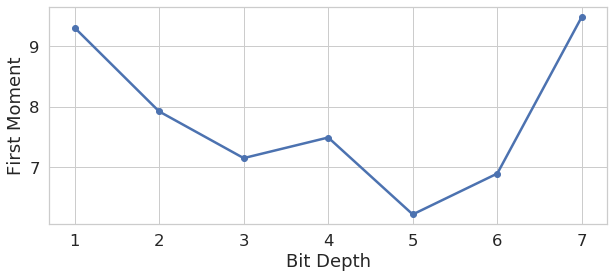}
        \caption{PAN Task}
        \label{fig:freq_dist_pan}
    \end{subfigure}
    \caption{Average spatial frequency of illuminated images for each bit-depth configuration}
    \label{fig:freq_dist}
\end{figure}

We are also able to compare the distribution of optimized LED patterns as a function of task difficulty (i.e., output label bit depth). Although the importance of establishing a learned pattern to improve network performance is clear (section \ref{sec:perf_vs_depth}), the trend among the spatial patterns themselves is less so. We found that, in general, the lower bit-depth patterns generated images which put more emphasis on higher frequency information, and vice versa.  Figure \ref{fig:freq_dist} shows how the average resolved spatial frequency power decreases with increased bit depth, although this trend was noisy and several bit-depths were exceptions. The values were determined by finding the first moment of each illuminated images' spatial frequency power representation along it's radius, more details are provided in Appendix \ref{app:freq}. We hope to examine more possible trends in future work. 

Although there is only a small amount of variation present in the overall system performance, the LED patterns themselves do vary across random seeds. Figure \ref{fig:all_patterns} shows examples and statistics of this variance across both tasks. We hypothesize that there are two factors driving this variance. First, images acquired from high-angle illumination (particularly dark-field) had lower intensity values on average and therefore contributed less to the synthesized image. This translates to a higher degree of allowed variance in the optimized result, since as the brightness value of an LED illuminating from a high angle doesn't impact the image as much as the more centrally located LEDs, and hence the gradients driving its value will be fairly weak. The use of the L1 norm penalty reduced this variance across both tasks. 

Second, there is clearly a coupling between the solution of the neural network and the parameterization of the LED array. The joint optimization procedure followed in this work uses this coupling to get higher performance results, and shows that higher performance is consistently obtained. But, as an adverse effect, it also decreases our ability to draw conclusions from the end parameterization of the LED array. This is because when a deep neural network is being trained, it can arrive at one of many local minima (finding the global minima is highly unlikely). Due to the network arriving and terminating at a local minima, the LED parameterization is such that the lighting is optimized \textit{for that specific local minima}, instead of what might be globally optimal. Within section \ref{sec:reg_impact}, we comment on one method of alleviating the impact of arriving at local optima through the use of targeted regularization.

\subsection{Impact of Regularization on LED Parameterization}
\label{sec:reg_impact}
Many of the trials presented here were run both with and without the addition of regularization on the weights representing the illumination pattern. Upon comparison, we found that the addition of this regularization did not impact our target performance statistic (MSE on the test set), however it did have a large effect on the LED patterns themselves. Most notably, when the LED weights are left unconstrained, the network tends towards an illumination pattern configuration using many LEDs at high dark-field angles. When constrained to use fewer LEDs, the optimized pattern instead utilizes bright-field and dark-field LEDs at lower angles. Although information contributed from higher angles may still be useful overall, the L1 norm prioritizes illumination angles that consistently contribute large amounts of information. In this respect a standard L1 norm may not be ideal for this kind of optimization, investigation into a more suitable regularization strategy will be investigated in future work.

\begin{figure}[tb]
    \centering
    \begin{subfigure}[t]{\columnwidth}
        \centering
        \includegraphics[width=\textwidth]{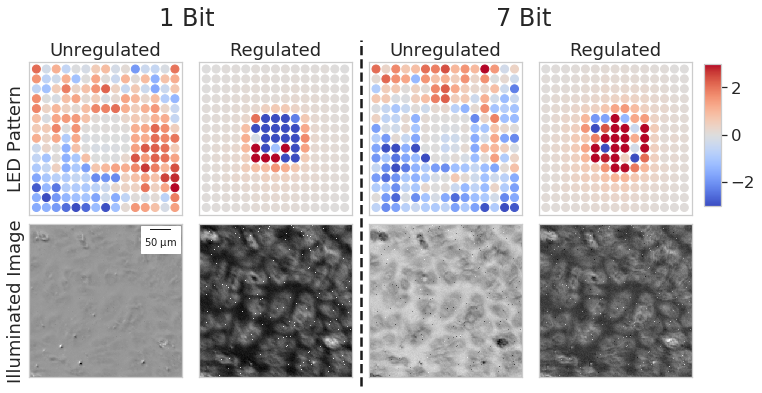}
        \caption{HeLa cell task LED configurations after training}
        \label{fig:hela_reg_compare}
    \end{subfigure}%
    \\
    \begin{subfigure}[t]{\columnwidth}
        \centering
        \includegraphics[width=\textwidth]{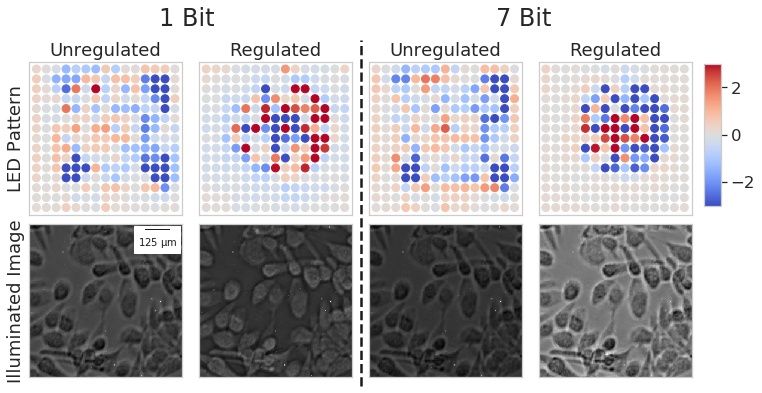}
        \caption{Pancreatic Cancer cell task LED configurations after training}
        \label{fig:pan_reg_compare}
    \end{subfigure}
    \caption{Visualization of the impact of regularization on the optimized LED patterns, and corresponding illuminated image. Examples shown across both tasks and a subset of bit-depths. Illuminated images contain both positive and negative numbers and are auto-scaled for maximum contrast.}
    \label{fig:reg_no_reg}
\end{figure}

In Figure \ref{fig:reg_no_reg}, we show side-by-side comparisons for both cases across a representative set of bit depths. This comparison clearly shows that not only do the regularized patterns contain less within-trial spatial variance, but also they are potentially more spatially interpretable. We postulate that this is because the added regularization is having the desired effect of forcing the optimization to find a parameterization that is robust to perturbations (caused by the addition of noise) while simultaneously having the minimum overall energy (caused by the L1 penalty). While performance in the simulated environment remains the same (with and without regularization), we hypothesize that these spatially smoother LED patterns are more robust to changes in the eventual physical setup. However, we leave testing this hypothesis for future work.

\section{Conclusion}
In summary, we have presented a novel method for developing image capture systems that can be optimized for deep learning based image-to-image inferencing tasks. By placing the physical parameters of the microscope in the gradient pathway, we jointly optimized the way an image was captured with the way it was processed. This allows imaging systems to sample data, not based on what a human experimenter prefers, but governed through optimization.

The two experiments we performed show that this technique is robust in its improvement, with MSE under each configuration being minimized by the "learned" or jointly optimized approach. Furthermore, our experiments show that this technique provides performance improvement for even the simplest version of inference and this improvement increases with problem difficulty and illumination sensitivity.

Our experimental results show that the physical parameters of a microscope play an important role in deep learning image-to-image inference systems. By allowing the joint optimization of illumination and image processing we achieve consistently better performance than all tested alternatives. We hope our results continue to motivate the imaging and machine learning community to re-examine how they capture data and continue to develop understanding of the connection between data capture and data processing.

\section{Acknowledgements}
We would like to thank Jaebum Chung for collecting and providing the data used for the \textit{HeLa} task, as well as Syliva Ceballos and William Eldridge for preparing the Pan 16 samples. Research reported in this publication was supported by the National Institute Of Neurological Disorders And
Stroke of the National Institutes of Health under award number RF1NS113287, as well as the Duke-Coulter Translational Partnership.
\bibliographystyle{ieeetr}
\bibliography{refs}
\appendices
\section{Illuminated Image Frequency Analysis}
\label{app:freq}
\begin{figure*}[tb]
    \centering
    \includegraphics[width=2\columnwidth]{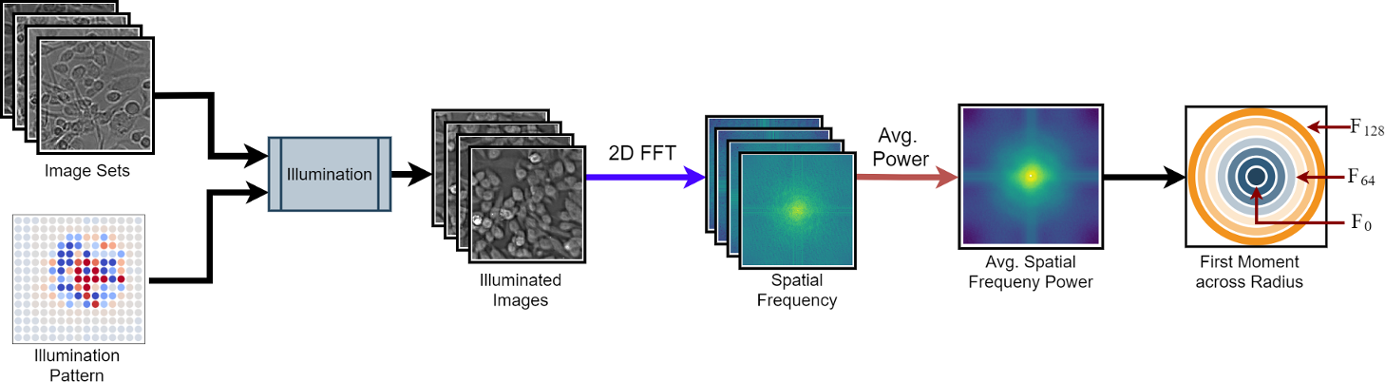}
    \caption{Process for determining the average resolved spatial frequency for a particular illumination pattern and image set.}
    \label{fig:spatial_freq}
\end{figure*}

The illumination patterns shown in this work offer a limited amount of insight into how the DNN affects image formation. To develop a better understanding of this process, we examined the illuminated images themselves, and examined how they change as we manipulated the difficulty of each inference task. To do this we define a metric called \textit{Average Spatial Frequency Power} which summarizes the spatial frequency of an image. Figure \ref{fig:spatial_freq} shows a diagram illustrating the process of computing this metric. 

First, the images from a representative sample of the dataset (in this case, the test set) are illuminated with the optimized LED pattern, creating a set of illuminated images. Each of these images is then put through a 2D Fourier transform (2D FFT) to establish a spatial frequency domain representation, which we squared (i.e., element-wise square) to form a per-image power spectrum. We then computed the average power spectrum across all representative images per sample. Finally the average spatial frequency power is calculated by taking the first moment of the frequency power representation along the radius of the spatial frequency distribution, resulting in the 1D plot shown in the main text.

\section{Neural Network Configuration Parameters}
\label{app:nn_config}
Throughout all experiments the same neural network architecture was used. We detail the key parameters of our U-Net architecture in Table \ref{tab:network_architecture}. Within Table \ref{tab:network_params} we report the hyperparameters used during training. With the exception of the noise level these hyperparameters were the same across tasks.

\begin{table}[htb]
    \centering
    \caption{Neural Network Architecture}
    \begin{tabular}{||c|c|c||}
    \hline
        Parameter & Value \\
        \hline
        \hline
        Initial Filters & $16$\\ 
        \hline
        Filter Expansion Ratio & $2$\\
        \hline
        Convolutional Layers per Block & $2$\\
        \hline
        Convolutional Down-sampling Blocks & $5$ \\
        \hline
        Convolutional Up-sampling Blocks & $5$ \\
        \hline
        Convolutional Kernel Size & $3\times3$\\
        \hline
        Activation Function & ReLU \\
        \hline
        Final Activation Function & Sigmoid \\
        \hline
        BatchNorm Frequency  & After every convolution\\
        \hline
        Convolutional Padding Amount & (1,1) zeros\\
        \hline
    \end{tabular}
    \label{tab:network_architecture}
\end{table}
\begin{table}[htb]
    \centering
    \caption{Training Hyperparameters}
    \begin{tabular}{||c|c|c||}
    \hline
        Hyper Parameters & HeLa Task & PAN Task\\
        \hline
        \hline
        Optimizer & Adam & Adam\\ 
        \hline
        Initial Learning Rate & $0.005$ & $0.005$\\
        \hline
        LR Reduction Factor & $\sqrt{10}$& $\sqrt{10}$\\
        \hline
        LR Reduction Patience & 5 & 5\\
        \hline
        Noise Level (k) & 0.1 & 0.3 \\
        \hline
        L1 Penalty & 0.0004 & 0.0004\\
        \hline
        Batch Size & 4 & 4\\
        \hline
    \end{tabular}
    \label{tab:network_params}
\end{table}
\section{Additional LED Patterns and Illumination Examples}
\label{app:patterns}
\begin{figure*}[bh]
    \centering
    \begin{subfigure}[t]{\size\columnwidth}
        \centering
        \includegraphics[width=\textwidth]{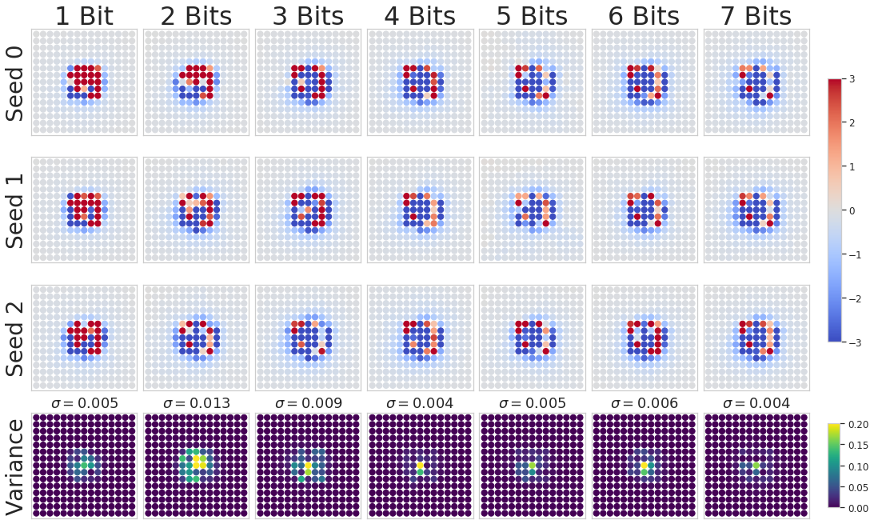}
        \caption{HeLa Task - Red LED Patterns}
        \label{fig:hela_red}
    \end{subfigure}
    ~
    \begin{subfigure}[t]{\size\columnwidth}
        \centering
        \includegraphics[width=\textwidth]{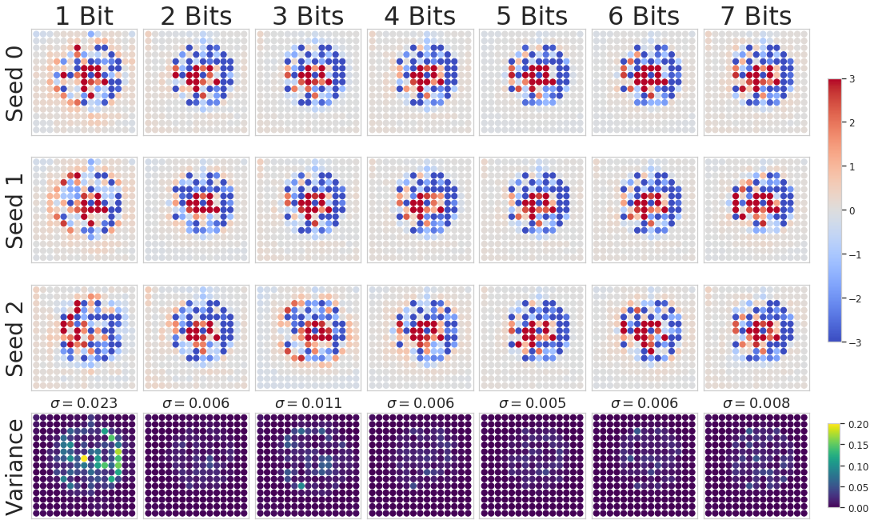}
        \caption{PAN Task - Red LED Patterns}
        \label{fig:pan_red}
    \end{subfigure}
    \\
    \begin{subfigure}[t]{\size\columnwidth}
        \centering
        \includegraphics[width=\textwidth]{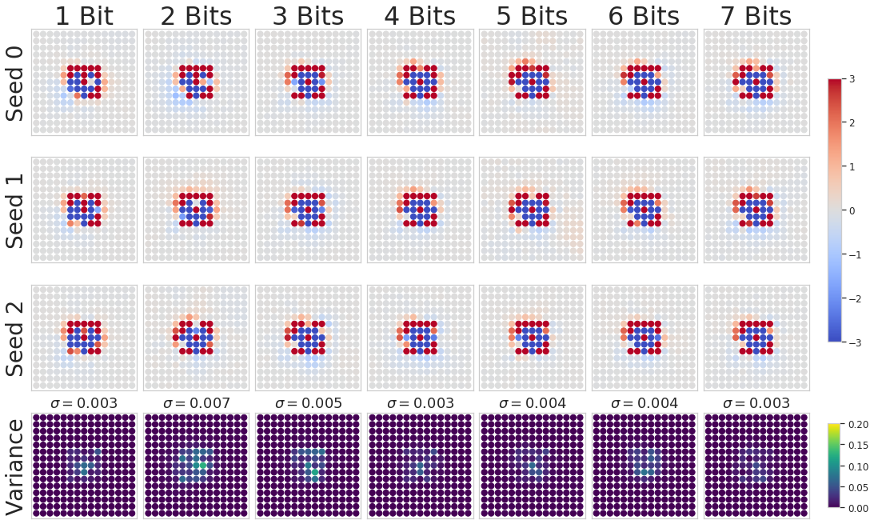}
        \caption{HeLa Task - Green LED Patterns}
        \label{fig:hela_green}
    \end{subfigure}
    ~
    \begin{subfigure}[t]{\size\columnwidth}
        \centering
        \includegraphics[width=\textwidth]{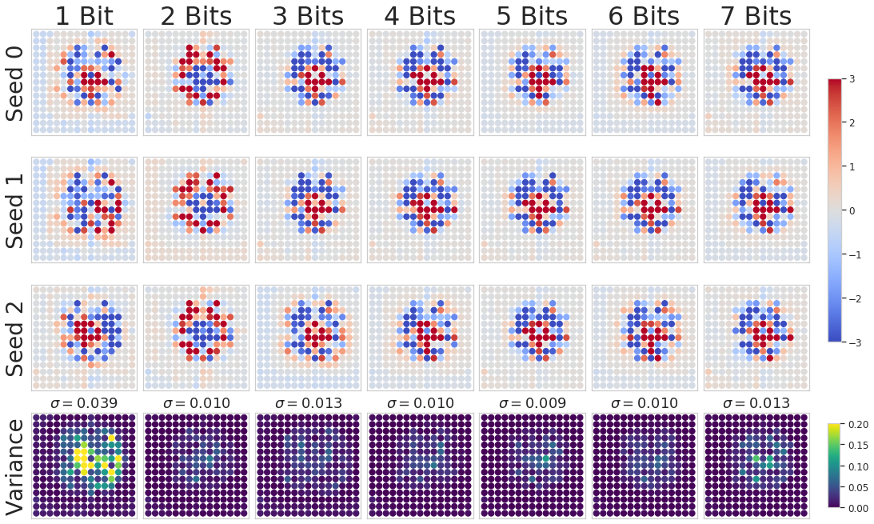}
        \caption{PAN Task - Green LED Patterns}
        \label{fig:pan_green}
    \end{subfigure}
    \\
    \begin{subfigure}[t]{\size\columnwidth}
        \centering
        \includegraphics[width=\textwidth]{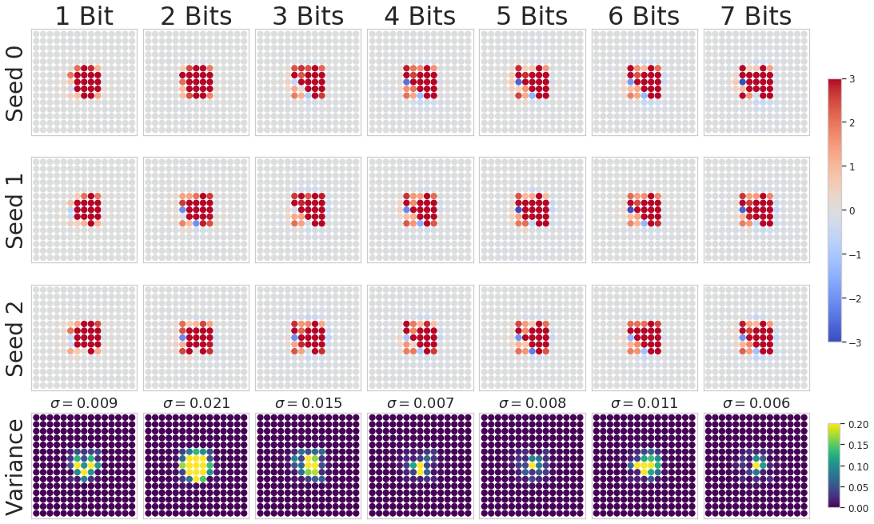}
        \caption{HeLa Task - Blue LED Patterns}
        \label{fig:hela_blue}
    \end{subfigure}
    ~
    \begin{subfigure}[t]{\size\columnwidth}
        \centering
        \includegraphics[width=\textwidth]{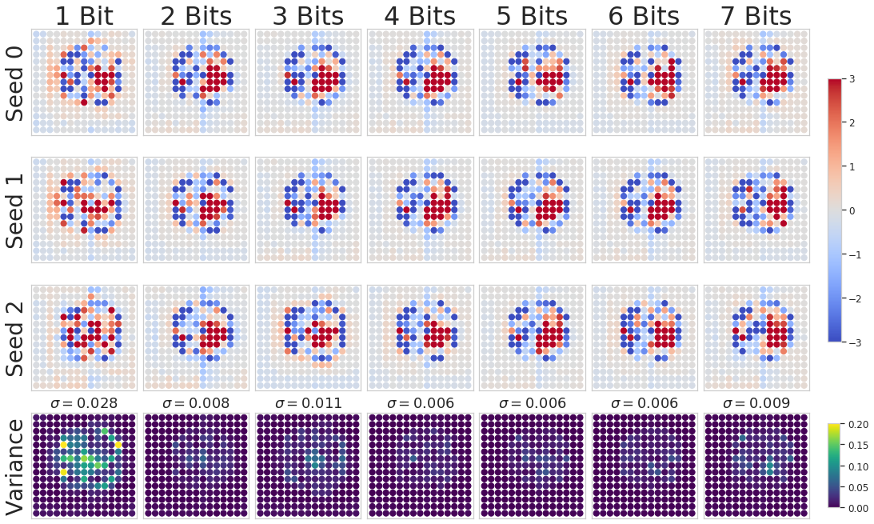}
        \caption{PAN Task - Blue LED Patterns}
        \label{fig:pan_blue}
    \end{subfigure}
    \caption{All LED patterns across both tasks and all bit-depths. The sign of each pattern is adjusted on a per-run basis to allow for easier interpretation.}
    \label{fig:all_patterns}
\end{figure*}

To supplement the LED patterns shown in the main text Figure \ref{fig:all_patterns} shows every LED pattern across three random seeds on a per-color basis. The variance for each individual LED is also plotted to illustrate how random seeds effect converged  LED patterns. In general we observe that the average variance across the patterns decreases with bit-depth.

\begin{figure*}[tb]
    \centering
    \begin{subfigure}[t]{\columnwidth}
        \centering
        \includegraphics[width=0.9\textwidth]{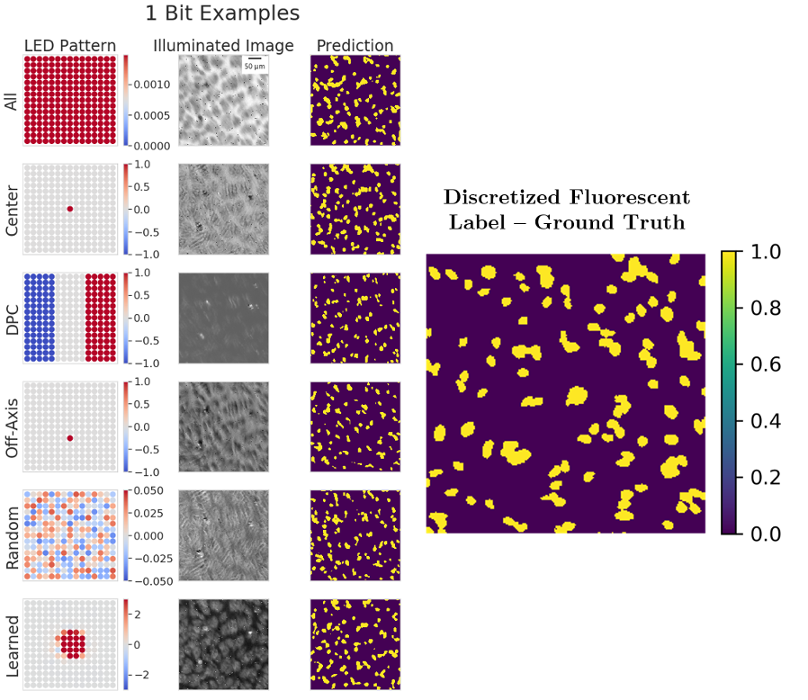}
        \caption{HeLa Task - One Bit Configuration}
        \label{fig:hela_example_2}
    \end{subfigure}%
    ~
    \begin{subfigure}[t]{\columnwidth}
        \centering
        \includegraphics[width=0.9\textwidth]{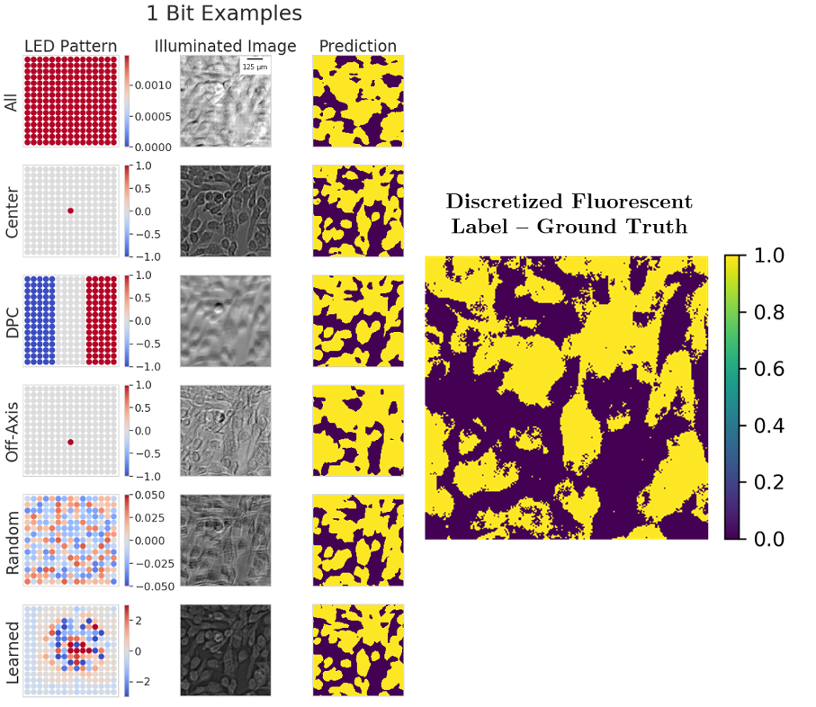}
        \caption{PAN Task - One Bit Configuration}
        \label{fig:pan_example_2}
    \end{subfigure}%
    \\
    \begin{subfigure}[t]{\columnwidth}
        \centering
        \includegraphics[width=0.9\textwidth]{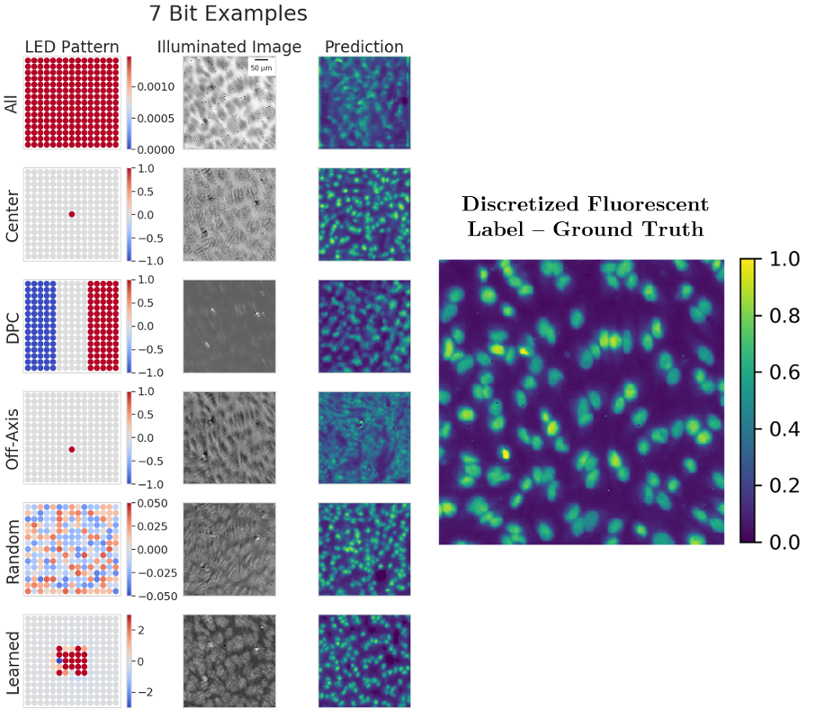}
        \caption{HeLa Task - Seven Bit Configuration}
        \label{fig:hela_example_7}
    \end{subfigure}
    ~ 
    \begin{subfigure}[t]{\columnwidth}
        \centering
        \includegraphics[width=0.9\textwidth]{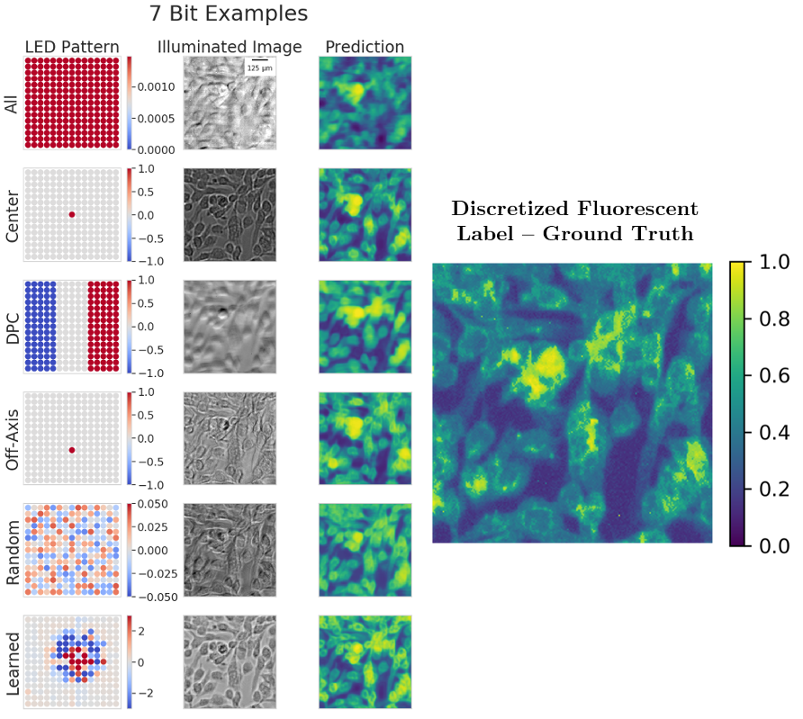}
        \caption{PAN Task - Seven Bit Configuration}
        \label{fig:pan_example_7}
    \end{subfigure}
    \caption{Examples for a sub-set of bit depths of illumination patterns, corresponding illuminated images, and inference results. Across all tasks and bit-depth configurations the learned pattern outperforms the alternatives.}
    \label{fig:examples}
\end{figure*}

To highlight the differences in the final illuminated image, as well as the impact on the inference results Figure \ref{fig:examples} shows examples for both tasks across a sub-set of bit depths. The learned (DNN-optimized) patterns consistently out-perform the alternatives, yielding inference results which are closer to the discretized fluorescent label.

\newpage

\end{document}